\documentclass[12pt]{article}

\usepackage{setspace,graphicx,epstopdf,amsmath,amsfonts,amssymb,amsthm,versionPO}
\usepackage{marginnote,datetime,enumitem,subfigure,rotating,fancyvrb}
\usepackage{hyperref,float}
\usepackage[longnamesfirst]{natbib}
\usdate
\usepackage{xltabular}
\usepackage{caption}
\usepackage{booktabs}
\usepackage{xtab}
\usepackage{lscape}

\excludeversion{notes}		
\includeversion{links}          

\iflinks{}{\hypersetup{draft=true}}

\ifnotes{%
\usepackage[margin=1in,paperwidth=10in,right=2.5in]{geometry}%
\usepackage[textwidth=1.4in,shadow,colorinlistoftodos]{todonotes}%
}{%
\usepackage[margin=1in]{geometry}%
\usepackage[disable]{todonotes}%
}

\let\oldmarginpar\marginpar

\makeatletter\let\chapter\@undefined\makeatother 

\newcommand{\smalltodo}[2][] {\todo[caption={#2}, size=\scriptsize, fancyline, #1] {\begin{spacing}{.5}#2\end{spacing}}}
\newcommand{\rhs}[2][]{\smalltodo[color=green!30,#1]{{\bf RS:} #2}}

\newcommand{\rhsfn}[2][]{
\renewcommand{\marginpar}{\marginnote}%
\smalltodo[color=green!30,#1]{{\bf RS:} #2}%
\renewcommand{\marginpar}{\oldmarginpar}}

\usepackage{indentfirst} 
\usepackage{jfe}          

\begin{document}

\setlist{noitemsep}  


\title{The Effect of Antitrust Enforcement on Venture Capital Investments \footnotetext{$\dagger$ Simon Business School, University of Rochester, E-mail: wentian.zhang@simon.rochester.edu.}}

\author{Wentian Zhang $^{\dagger}$}

\date{\today}              


\renewcommand{\thefootnote}{\fnsymbol{footnote}}

\singlespacing

\maketitle

\vspace{-.2in}
\begin{abstract}
\noindent This paper studies the effect of antitrust enforcement on venture capital (VC) investments and VC-backed companies. To establish causality, I exploit the DOJ's decision to close several antitrust field offices in 2013, which reduced the antitrust enforcement in areas near the closed offices. I find that the reduction in antitrust enforcement causes a significant decrease in VC investments in startups located in the affected areas. Furthermore, these affected VC-backed startups exhibit a reduced likelihood of successful exits and diminished innovation performance. These negative results are mainly driven by startups in concentrated industries, where incumbents tend to engage in anticompetitive behaviors more frequently. To mitigate the adverse effect, startups should innovate more to differentiate their products. This paper sheds light on the importance of local antitrust enforcement in fostering competition and innovation.

\end{abstract}



\thispagestyle{empty}

\clearpage

\doublespacing
\setcounter{footnote}{0}
\renewcommand{\thefootnote}{\arabic{footnote}}
\setcounter{page}{1}

\section{Introduction}
Venture capital (VC) and startups backed by VCs are important contributors to the economy. According to \cite{gornall2021economic}, around half of the U.S. public firms founded within the past fifty years were financed by a VC fund in their early stage, accounting for 41\% of the total market capitalization. Moreover, the macroeconomic impact of VC investments is much greater than that indicated by market valuations since the profound contribution of innovations generated by VC-backed startups is not fully reflected in the valuations \citep{opp2019venture}. However, these startups are vulnerable to the anticompetitive conduct of incumbents (e.g., killer acquisition \citep{cunningham2021killer}, predatory or exclusionary acts, etc.). To promote competition in the product market and foster innovation, antitrust laws intend to protect startups from being abused and deterred by the monopoly power of incumbents. Recently, these protections are becoming more intense. In December 2022, the U.S. Federal Trade Commission (FTC) sued Microsoft's acquisition of Activision for harming the competition in gaming consoles and subscription services. Shortly thereafter, in January 2023, DOJ antitrust division filed its second antitrust lawsuit against Google in just over two years, seeking to break up Google’s online advertising business. These major antitrust cases brought against tech giants have caused debates among VCs on whether more intense antitrust enforcement could hurt the startups they are meant to protect \footnote{https://www.cnbc.com/2021/07/24/vcs-start-ups-will-suffer-from-antitrust-bills-targeting-big-tech.html}. Some VCs express the concern that intense antitrust enforcement will deter the incumbents from acquiring the startups and thus stifle an important pathway for VC exit. Other VCs suggest that adequate enforcement on anticompetitive practices can level the playing field. Therefore, it's worth investigating how antitrust enforcement affects VCs' investment decisions and VC-backed startups.

Antitrust laws in the U.S. play a crucial role in promoting competition and innovation by prohibiting anticompetitive mergers and business practices. Specifically, the antitrust laws regulate the three main aspects: cartels and anticompetitive agreements (Sherman Act, Section 1), anticompetitive business practices (e.g., exclusionary acts, predatory acts, etc.) used by dominant firms (Sherman Act, Section 2), and mergers that may significantly reduce competition (Clayton Act). While the antitrust laws undoubtedly restrain the market power of incumbents, their effects on new entrants and the VCs investing in these startups are mixed. On the one hand, antitrust enforcement protects the new entrants from being abused by the anticompetitive practices of incumbents (Sherman Act, Section 2) and thus increases startups' chance of survival. In this case, antitrust enforcement motivates the VCs to invest more in startups. On the other hand, intense antitrust enforcement may refrain VCs from investing in new entrants and thus harm the development of startups for the following three reasons. First, increasing antitrust enforcement on mergers (Clayton Act) deters the incumbents from acquiring new entrants, reducing the chance of a successful VC exit. Second, under intense enforcement (Sherman Act, Section 1), firms are less likely to form cartels that collude to raise product prices, leading to more vigorous competition, lower prices, and fewer profits for both incumbents and new entrants. Startups' poor performance in the product market makes VCs reluctant to continue investing in these companies. Third, strict enforcement on monopolization (Sherman Act, Section 2) makes the startups less likely to take advantage of their market power even after surviving the competition and gaining significant market share, leading to fewer investments by VCs.

To identify the causal effect of the antitrust enforcement on VC investments and VC-backed startups, I exploit the DOJ Antitrust Division field office closures in 2013 as an exogenous shock to the antitrust enforcement intensity in the local market. DOJ Antitrust Division closed 4 of its 7 field offices in 2013 in response to the call for a more efficient federal government by the Obama administration. The field offices mainly conduct investigations and litigation on anticompetitive business practices. They also handle merger cases depending on their available resources. The office closures significantly reduce the antitrust enforcement in the local market near the closed offices \citep{ha2024motivating}. Given this regulation change, I employ difference-in-differences method. The treated units are startups that become farther away from antitrust field offices after office closures in 2013 and thus whose local market is less policed by the antitrust division. 

To understand how VCs adjust investments to accommodate the reduction of antitrust enforcement in the local market of startups, I first examine the effect of antitrust office closures on the amount of VC investments in a startup. It turns out that the reduction in antitrust enforcement causes the amount of VC investments in startups in affected areas to decrease significantly. Specifically, the reduction in antitrust enforcement causes a 683.169 thousand dollar (17\%) decrease in the total amount invested in a startup within a year. I further explore whether the reduction in VC investment amount is driven by a smaller number of VC investors or less amount of investment by each investor. The result shows that both fewer investors per round of VC financing and less average amount of VC investment across investors contribute to the total investment reduction.

Since the ultimate goal of VC investors is to gain financial returns when they exit the startups, I next investigate the effect of antitrust enforcement on VC exits. A successful VC exit is defined as the startup being acquired or the startup going public \citep{sorensen2007smart, bottazzi2016importance, gu2022does}. I find that the lowered antitrust enforcement causes the likelihood of a successful VC exit to decline by 2.1 percentage points. I further decompose the VC exits into two categories and find that the smaller chance of success is mainly driven by less likelihood of exit by going public.

Given that startups lose investments from their major funding source, VCs, I further explore the subsequent effect on the startups' innovation. On the one hand, startups are motivated to innovate and differentiate their products from incumbents to avoid being abused by the incumbents' monopoly power. On the other hand, the startups receive less financial support from VCs, which may hinder them from investing more in risky R\&D projects. I measure the innovation output with the number of patents generated by startups and find that startups' innovation output is negatively affected by a reduction in antitrust enforcement. Specifically, the reduction in antitrust enforcement causes the number of patents applied by a startup within a year to reduce significantly by 12.5\% - 17.2\%. The negative effect is especially pronounced from the 4th year after the event and onwards. This result indicates a long-term negative effect on firms in areas less policed by the antitrust division.

In addition to the effect on startups already invested by VCs before the antitrust field office closures, I also examine the extensive margin, i.e., the effect on the number of new startups invested by VCs. I employ difference-in-differences method at the state level with the antitrust field office closures as an exogenous shock to the enforcement intensity within a state. The treated units are states overseen by the offices closed in 2013. Compared with the states overseen by the remaining three field offices, the number of new VC-backed startups in the treated states decreases by 11.8\% more. This result implies that VCs not only decrease their investments in affected startups already in their portfolios but also avoid investing in new startups in the states less protected by the antitrust division. 

I further explore the potential mechanisms driving the negative effect of the reduction in antitrust enforcement. First, I show that the negative effect is more prominent for startups in more concentrated industries. In more concentrated industries, anticompetitive practices of incumbents take place more often since these incumbents with large market shares have a deeper pocket of cash to predate the startups and more power to exclude the startups from the business. This result suggests that negative effects on VC-backed startups are driven by the potential anticompetitive practices of monopolists in more concentrated industries. 

Second, I find that innovative startups are less affected by the reduction in antitrust enforcement. This result indicates that VC investors are confident that innovative startups can differentiate their products from the incumbents' and thus are more likely to survive and thrive. In other words, the innovation ability of startups can mitigate the negative effect caused by less protection from the antitrust division.

In summary, this study finds that the reduction in antitrust enforcement refrains VC investors from investing in startups in less protected areas and subsequently harms the innovation performance of these startups. This is because startups in areas less policed by the antitrust division are likely to be abused by the market power of local incumbents. These results suggest that antitrust enforcement plays an important role in nurturing the development of new entrants to promote competition and innovation. This study highlights the importance of strengthening regional antitrust enforcement to motivate VC investments in startups, promote competition, and foster innovation.

This paper makes the following contributions. First, this paper contributes to the literature on venture capital. My study is especially related to prior VC literature studying the relationship between product market competition and VC investments. For example, \cite{hellmann2000interaction} document that VCs are more likely to invest in innovator startups rather than imitator startups. \cite{bayar2012drives} show that startups in more competitive markets tend to exit by going public instead of acquisition. \cite{cumming2022product} find that product market competition increases the likelihood of VC staged financing and causes more financing rounds. \cite{pham2023venture} find that VCs apply a financing strategy less contingent on startups' short-term performance when startups are faced with deep-pocketed incumbents. This paper complements this strand of literature in the following three aspects. First, this study evaluates the role of competition policy in protecting new entrants by suppressing the anticompetitive practices of incumbents, which is the opposite of cash motivating more anticompetitive practices as in \cite{pham2023venture}. Second, while previous literature focuses on the effect on VC financing strategies (e.g., staging, exit methods, types of startups to invest in, etc.), this paper focuses on the amount of VC investments and the subsequent effect on VC-backed startups' innovation outcome. Third, the variation of antitrust enforcement at the state level allows me to explore the extensive margin, i.e., the effect on VCs' choice of new startups to invest in, and thus to suggest the economic effect at the aggregate level.

Second, this paper contributes to the large literature on the economic effect of antitrust enforcement. One strand of this literature finds that antitrust enforcement can be harmful. For instance, \cite{sproul1993antitrust} finds that antitrust prosecution leads to an increase in prices charged by firms indicted for price fixing. \cite{crandall2002does} find that antitrust policy does not provide any benefit to consumers and may have lowered consumer welfare in some cases. \cite{bittlingmayer2000kapital} look into the antitrust enforcement action against Microsoft from 1991 to 1997 and find that the series of antitrust enforcement actions is associated with declines in the market value of firms in the computer industry. The other strand of antitrust literature supports that antitrust enforcement is beneficial to the economy. For example, \cite{callander2022novelty} find theoretical evidence that new entrants innovate less boldly when they pursue the profits of being acquired by incumbents, indicating a positive effect of a strict antitrust policy on spurring the novelty of innovation by startups. \cite{dasgupta2019anticollusion} document that more competition brought by stronger antitrust enforcement increases investments. \cite{babina2023antitrust} find that DOJ antitrust lawsuits permanently increase employment and business formation in the targeted states, leading to an increase in the quantity of output and a decrease in the price of output at the same time. My paper complements the antitrust literature by discovering VC investments as an important financial channel at the micro level contributing to the effect of antitrust enforcement on economic activities at the macro level.

My paper is not the only one that looks into the antitrust field office closures event in 2013. Another three papers also look into this event. Among them, two papers discover evidence of anticompetitive practices of local incumbents following the office closure event. Specifically, \cite{ha2024motivating} find that firms in areas affected by the antitrust field office closures motivate their CEO to collude with local rivals by increasing the sensitivity of executive compensation to the rivals' performance. \cite{oh2022antitrust} finds that firms strategically adjust their disclosure of merger deals according to the antitrust enforcement intensity. On the contrary to these papers, \cite{deng2023market} documents that firms in areas affected by the antitrust office closures lose some market power. While all these papers focus on the effect on incumbents, my study instead reveals the effect of this regulation change on the new entrants and the investment decisions of their investors. 

The rest of this paper is organized as follows. Section 2 introduces the data sources and sample construction; Section 3 describes the identification strategy; Section 4 outlines the main empirical results; Section 5 discusses the potential mechanisms driving the results; Section 6 presents the robustness tests; Section 7 concludes the paper.

\section{Data}
My main data source is the VentureXpert database. I include VC-backed startups that received at least one round of VC funding during the period from 2008 to 2017. I further require the VC-backed startups to satisfy the following restrictions: first, the startups are headquartered in the U.S.; second, the startups are financed by VC in the U.S.; third, the startups are founded before Antitrust Division field office closures in 2013; fourth, the startups exist as a private firm for at least 1 year after 2013. I track each startup from 2008 to 2017 and exclude years when the startup has not been founded or has already exited, i.e., has gone IPO or has been acquired. The main sample is a startup-year level panel including 10444 startups during the period from 2008 to 2017.

I also employ other data sources, including PatentsView for patenting data and Compustat for the industry concentration measure.

\section{Identification Strategy}
To evaluate the effect of antitrust enforcement on VC investments and VC-backed startups, I exploit the DOJ Antitrust field office closures in 2013 as a plausibly exogenous reduction in the strength of antitrust enforcement following \cite{ha2024motivating, oh2022antitrust, deng2023market}. This regulatory change makes the anticompetitive conduct by local incumbents in the affected areas harder to detect and thus exposes the startups in the affected areas to more risks of being abused by the enlarging market power of the incumbents. I discuss the background of this regulation, the functions of antitrust division field offices, and the main specification in this section.

\subsection{DOJ Antitrust Office Closures}
DOJ Antitrust Division's mission is to promote competition by enforcing antitrust laws. The antitrust division's field offices mainly conduct criminal investigations and litigation on anticompetitive conduct, e.g., cartels, bid rigging, exclusionary acts, predatory acts, etc. They also handle civil merger and nonmerger cases if resources are available (DOJ Antitrust Division, 2001). In 2013, DOJ Antitrust Division closed 4 of its 7 field offices in Atlanta, Cleveland, Dallas, and Philadelphia. Some of the attorneys and supporting staff in the closed offices were relocated to the remaining offices in Chicago, New York, San Francisco, and the division headquarters in Washington, D.C. This regulatory change is a part of the efforts by the Obama Administration to make the federal government more efficient. Since this regulation is due to the government's budget cut, the change in the strength of antitrust enforcement is plausibly exogenous to the economic environment faced by the individual companies.

The reduction effect of field office closures on antitrust enforcement is significant. Antitrust filings in the 23 affected states and territories decline dramatically after 2013, while the filings in the unaffected states do not change in the same way \citep{ha2024motivating}. Possible reasons for the reduction in the strength of antitrust enforcement are as follows. First, since the firms in the affected areas are farther away from the field offices, the costs for the attorney to travel from the office to the affected area increase. Second, attorneys familiar with the local legal system quit their jobs even though they were offered to relocate to the remaining offices (The Washington Post, 2011). According to an attorney at the Atlanta office of Antitrust Division, "San Francisco isn't going to go down to Texas to work on a case ... And someone from New York can't go down to a grand jury in Dallas and prevail. They just don't know the people and know how to do the cases" (The Washington Post, 2011). 



\subsection{Main Specification}
I employ difference-in-differences method by exploiting the Antitrust Division field office closures in 2013 as an exogenous change in the strength of antitrust enforcement following \cite{ha2024motivating, oh2022antitrust, deng2023market}. Startups becoming farther away from the antitrust field offices after the office closures in 2013 are the treated group, while startups whose distance to the antitrust field offices remains unchanged are the control group. This is because the local incumbents near startups farther away from the field offices are less policed by the antitrust division and are more likely to harm the startups by engaging in anticompetitive practices. To distinguish treated firms from control firms, I first measure the increase in geographical distance from the headquarter of startup $i$ to the closest DOJ antitrust field office before and after the office closures, $\Delta Distance_{i}$. Then if $\Delta Distance_{i} > 0$, firm $i$ is in the treated group; if $\Delta Distance_{i} = 0$, firm $i$ is in the control group.

To evaluate the causal effect of the antitrust enforcement on VC investments and VC-backed startups, the main specification is a startup-year level regression:

\begin{equation}
    Y_{it} = \alpha + \beta Treat_{i} \times Post_t + \theta_i + \gamma_t + \epsilon_{it},
\end{equation}
where the outcome variable $Y_{it}$ can be one of the following: the total amount of VC disbursement for company $i$ in year $t$; the average amount of VC disbursement per investor for company $i$ in year $t$; the average number of investors per round of VC investments in company $i$ in year $t$. The $Post_t$ is a dummy variable that equals 1 in and after 2013 when antitrust field offices closed. I include startup fixed effect to control for time-invariant startup characteristics that could confound the VC investments and the strength of antitrust enforcement, e.g., the startup quality, sector, location, financing stage when the event occurred, etc. I also include year fixed effect to control for the time trend and events affecting all startups across years, e.g., the economic cycle. The coefficient of interest, $\beta$, measures the causal effect of the reduction in antitrust enforcement on VC investments.

Since VC investments in a startup only occur once in a few years, the distributions of outcome variables are right-skewed with masses of values at zero. This distributional feature makes linear regression inefficient. Therefore, in addition to the previous linear model, I use Poisson regression to estimate the effect on VC investments \citep{cohn2022count}.



\section{Empirical Results}

\subsection{Summary Statistics}
Table \ref{table:summary_stats} presents the summary statistics of the main sample used for the analysis in this paper. The treatment variable, $Treat$, indicates startups becoming farther away from the antitrust field offices after the office closures in 2013. The mean of the treatment variable is 0.279, implying that around 28\% of startups in the sample are in the treated group. $\Delta Distance$ measures the change in the distance from the startup's headquarter to the closest field office before and after 2013. The average change in the distance is 80.289 miles. As for the main outcome variable, the average amount of VC investments in a startup within a year is about 2,538,258 dollars. However, the median of the amount of VC investments is 0, indicating a skewed distribution concentrated at 0. This is also true for the amount per investor and the number of investors involved per round. The Herfindahl index (HHI) measures the market concentration of public companies within the startup's industry. The larger the HHI is, the more concentrated the industry is. The HHI ranges from 0.059 to 0.999 and the average is 0.256. These statistics indicate that the startups span across industries, ranging from highly competitive industries to highly concentrated industries.

\subsection{VC Investments}
I first investigate the effect of antitrust enforcement on VC investments in VC-backed startups. Table \ref{table:vc_inves} presents the estimation results of the DID model described in the previous section. The outcome variable of the first two columns is the total amount of VC disbursement in a startup within a year. Column (1) reports the OLS estimation result for equation (1). The coefficient on the interaction term is significantly negative, indicating that VC invests less in startups in the areas with less antitrust enforcement. Specifically, the reduction in antitrust enforcement causes VC investments in a startup in the area less policed by the antitrust division to decrease by 683.169 thousand dollars within a year. The decrease in VC investments accounts for 27\%  (= 683.169/2538.258) of the average VC investments in a startup within a year. Similarly, the result of Poisson regression in column (2) shows that less antitrust enforcement causes VC investments in a startup to decrease by 17\% ($=e^{-0.186}-1$).

To further examine how the VC investors react to the antitrust policy change, I look into the effect on the number of investors per round of VC investments and the average amount of VC disbursement for a startup within a year across VC investors. The result of column (3) shows that the number of VC investors involved in each round of VC investments decreases by 7.7\% ($=e^{-0.08}-1$).  According to the result in column (4), the amount of VC disbursement per investor for startups in affected areas declines by 13.5\% ($=e^{-0.145}-1$). 

These results imply that VCs invest less in startups in the areas with less antitrust enforcement, both by decreasing the investment amount and by avoiding investing in these startups at all.

\subsection{VC Investment Outcomes}
VCs' ultimate goal is to earn financial returns when they exit the startup. Thus, I examine the effect of antitrust field office closures on VC exits. According to the VC literature (e.g., \cite{sorensen2007smart, bottazzi2016importance, gu2022does}), IPO and M\&A are the two ways of a successful VC exit. I first estimate the effect on the probability of a successful VC exit, and then look into the effect on the probability of exiting by going public and exiting by acquisition, respectively. I run the following cross-sectional regression:

\begin{equation}
    Y_i = \alpha + \beta_1 Treat_i + \beta_2 Controls_i + \gamma_f + \epsilon_{i},
\end{equation}

where the outcome variable is a dummy variable equal to 1 if startup $i$ successfully exited after field office closures in 2013, and 0 otherwise; $Treat_i$ is a dummy variable equal to 1 if startup $i$ became farther away from the antitrust field offices since 2013; control variables include the log of the cumulative amount of VC investments received by startup $i$ up to the year before office closures (year 2012), startup age in 2012, and the log of real GDP per capita in 2012 of the state where the startup is located. I also control industry fixed effects ($\gamma_f$) to account for any unobserved industry characteristics potentially driving both the VC exit probability and the decision of antitrust field office closures. The coefficient of interest is $\beta_1$, which measures the difference between the exit probability of startups in areas less policed by antitrust field offices and that of startups in unaffected areas.

Table \ref{table:vc_exit} reports the estimated effect of antitrust field office closures on VC exits. According to the result in column (1), the reduction in antitrust enforcement causes the chance of a successful exit to decrease significantly. Specifically, the probability of a successful exit for startups in affected areas decreases by 2.1 percentage points. When I decompose the two pathways of a successful VC exit, I find that the decrease in the chance of a successful exit is mainly driven by a lower probability of the startup going public. For startups becoming farther away from the antitrust field offices, the probability of going IPO drops by 1.5 percentage points. In comparison, the probability of the startup being acquired decreases by 0.5 percentage points but the decrease is not statistically significant.

\subsection{Innovation of VC-Backed Startups}
Since VC-backed startups are important contributors to innovation in the U.S., I further examine the effect of antitrust field office closures on the subsequent innovation outcomes of VC-backed startups. Less antitrust enforcement could have both positive and negative effects on the innovation of VC-backed startups. On one hand, less antitrust enforcement leaves the startups threatened by the enlarging market power of local incumbents. Thus, to survive the product market competition, the startups will innovate more to differentiate their products from the incumbents'. In this case, we should see more innovation outcomes by the VC-backed startups. On the other hand, since the startups receive fewer investments from VCs according to the results in section 4.2, startups will refrain from costly and risky R\&D investments, and thus their innovation outcomes will decrease. The empirical result presented in this section answers how the innovation of VC-backed startups changes after the antitrust enforcement reduces.

I measure the innovation outcome with the number of patents applied by startup $i$ in year $t$ and use the Poisson regression in equation (1) to estimate the effect of antitrust office closures on innovation because the distribution of the number of patents is skewed and concentrated at 0. Since the time needed from an R\&D project initiation to patent application varies across industries, I also lag the number of patents by 1 - 4 years and repeat the estimation. The estimation results are presented in Table \ref{table:vc_numpat}. The reduction in antitrust enforcement causes the number of patents by a VC-backed startup to decrease significantly by 12.5\% - 17.2\%. This result supports the argument that startups refrain from investing in R\&D after receiving fewer investments from VCs and implies a negative real effect on startups in addition to the detrimental effect on investments.

\subsection{The Number of New VC-Backed Startups}
The previous sections investigate the effect of antitrust enforcement on VC investments on an intensive margin, i.e., the effect on startups already invested by VCs. In this section, I explore the effect on VC investments on an extensive margin, i.e., the effect on the number of new VC-backed startups.

I employ difference-in-differences method in which the treated states were overseen by the antitrust field offices that closed in 2013. Specifically, I run the following state-year level regression:
\begin{equation}
    Y_{st} = \alpha + \beta_1 Treat_{s} \times Post_{t} + \beta_2 Treat_{s} + \beta_3 Post_{t} + \epsilon_{st}, 
\end{equation}
where the outcome variable $Y_{st}$ is the number of startups receiving their first VC disbursement in state $s$ during year $t$. $Treat_{s}$ is a dummy variable equal to 1 if state $s$ was overseen by an antitrust field office closed in 2013. $Post_{t}$ is a dummy variable equal to 1 if the antitrust field offices have closed in year $t$. To control for unobservable time-invariant state features and common time trends across states that may confound the event, I also run an alternative regression where I control state fixed effects and year fixed effects instead of $Treat_{s}$ and $Post_{t}$.

Table \ref{table:new_vc_backed_companies} presents the estimation results of equation (2). The antitrust office closures cause the number of new VC-backed startups in affected states to decrease by 11.8\% (=$e^{-0.125}-1$). This result remains the same when I control for state fixed effects and year fixed effects. The result is also robust when I additionally control for the real annual GDP per capita within each state.

These results imply that VC investors avoid investing in new startups in areas becoming less policed by the antitrust division.

\section{Mechanisms}
This section investigates the potential mechanisms driving VCs' decision to decrease their investments in startups in areas less policed by the antitrust division. I first examine whether the potential anticompetitive conduct of incumbents drives this result by comparing the effect on startups in competitive industries with that on startups in concentrated industries. Then I examine if startups' ability to innovate and differentiate their products can mitigate the negative effect of less antitrust enforcement.

\subsection{Product Market Anticompetitive Conduct of Incumbents in Concentrated Industries}
After the antitrust field office closures in 2013, startups in the affected areas are less protected by the antitrust division. VCs will expect the startups in these areas to be more likely to suffer from the anticompetitive conduct of the monopolies. Given that new entrants in more concentrated industries are more likely to be abused by monopolization, we should observe that the startups in more concentrated industries are affected more negatively if the risk of being abused by the anticompetitive conduct indeed drives VCs investment decisions.

To test this mechanism, I divide the main sample into two subgroups. One group includes startups in more concentrated industries and the other group includes startups in more competitive industries. Industry concentration is measured by the Herfindahl-Hirschman index (HHI) of public firms in the startup's industry identified by the 4-digit SIC code. I repeat the analysis in equation (1) in each subgroup. According to the results in Table \ref{table:vc_inves_hhi}, the negative effects of the lowered antitrust enforcement are mainly driven by startups in more concentrated industries. Specifically, the total VC investments in a startup of more concentrated industries decrease by 22.1\% (1176.11 thousand dollars). In addition, the number of investors involved in each financing round reduces by 16.8\% and the amount invested per investor decreases by 18.3\%. In contrast, the effects on startups in more competitive industries are not significant. Thus, the reduction in VC investments is mainly driven by startups in more concentrated industries, indicating that potentially more frequent anticompetitive business practices by monopolists in concentrated industries is a plausible mechanism driving the result.

\subsection{The Startup's Potential to be a Future Monopolist}
While the reduction in antitrust enforcement leaves the VC-backed startups under threat of anticompetitive practices by monopolists, it's possible that VC investors expect these startups to take advantage of the loosened enforcement and be future monopolists after they survive the competition with the incumbents and other new entrants. Since innovative startups are able to differentiate their products and escape the competition, they are more likely to survive and gain monopoly power in the future. Thus, I examine whether VC investors' expectations for future monopoly is a mechanism mitigating the negative effect of less antitrust enforcement by comparing the effects on innovative startups and less innovative startups. If this mechanism is valid, then we should observe that the effect is more negative for less innovative startups.

I divide the VC-backed startups in the main sample into two subgroups by their average number of patents applied per year before the event year. This is because innovators tend to obtain more patents than imitators \citep{hellmann2000interaction}. Then I repeat the analysis in equation (1) for each subgroup. As shown in Table \ref{table:vc_inves_innovative}, less innovative startups are more negatively affected by the reduction in antitrust enforcement. Specifically, the lowered antitrust enforcement causes the total amount of VC investments in a less innovative startup to decrease by 21.3\% (712.221 thousand dollars). Additionally, the average amount invested per investor decreases by 16.6\%. These results imply that the ability to innovate and differentiate the product mitigates the negative effect of the reduction in antitrust enforcement, and this regulation change mainly hurts less innovative startups.

\section{Robustness Tests}

\subsection{Parallel Trend Tests}
\begin{figure}[!htb]
\centerline{\includegraphics[width=5in]{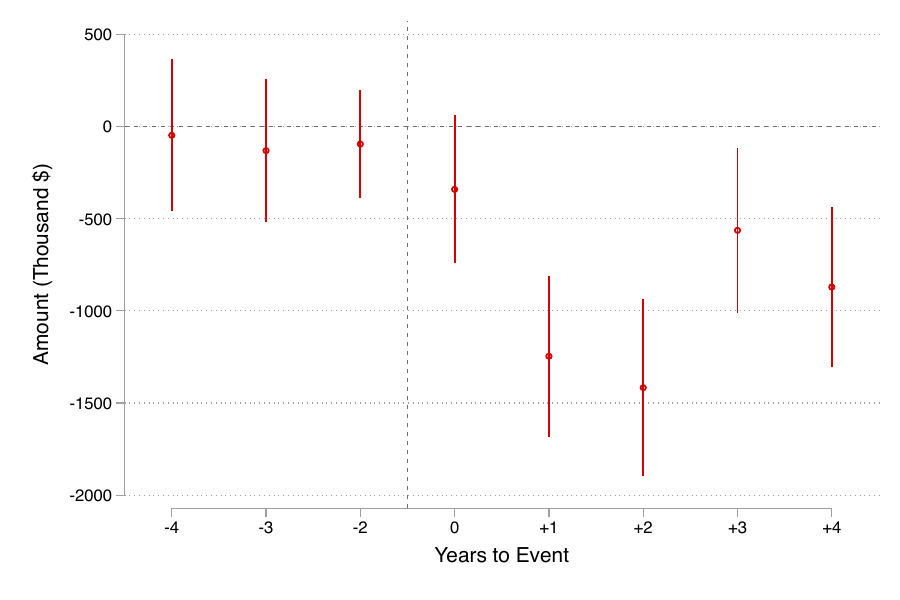}}
\caption{Dynamic Effect on the Amount of VC Investments (OLS)} 
\label{fig:vc_tot_amount_ols_parallel}
\end{figure}

\begin{figure}[!htb]
\centerline{\includegraphics[width=5in]{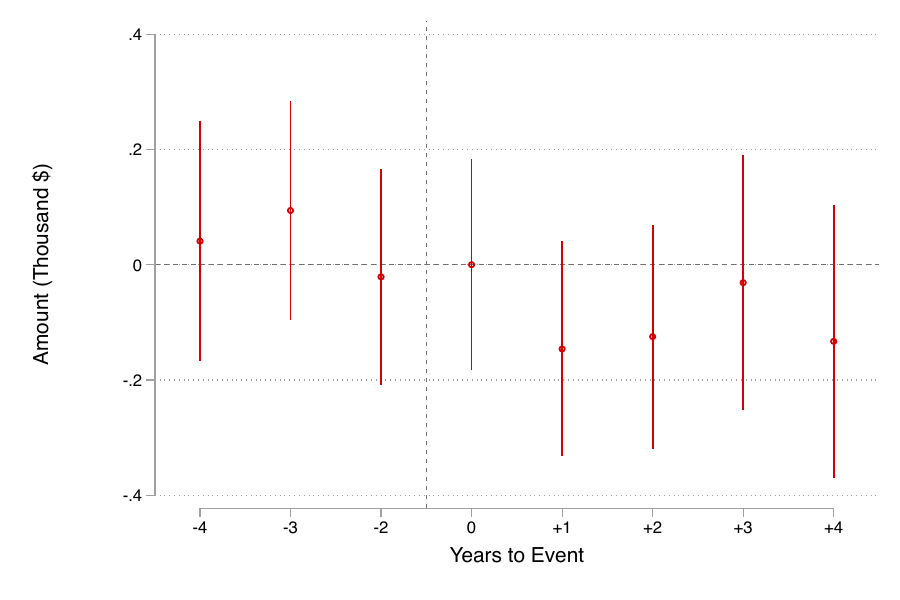}}
\caption{Dynamic Effect on the Amount of VC Investments (Poisson)} 
\label{fig:vc_tot_amount_poisson_parallel}
\end{figure}

\begin{figure}[!htb]
\centerline{\includegraphics[width=5in]{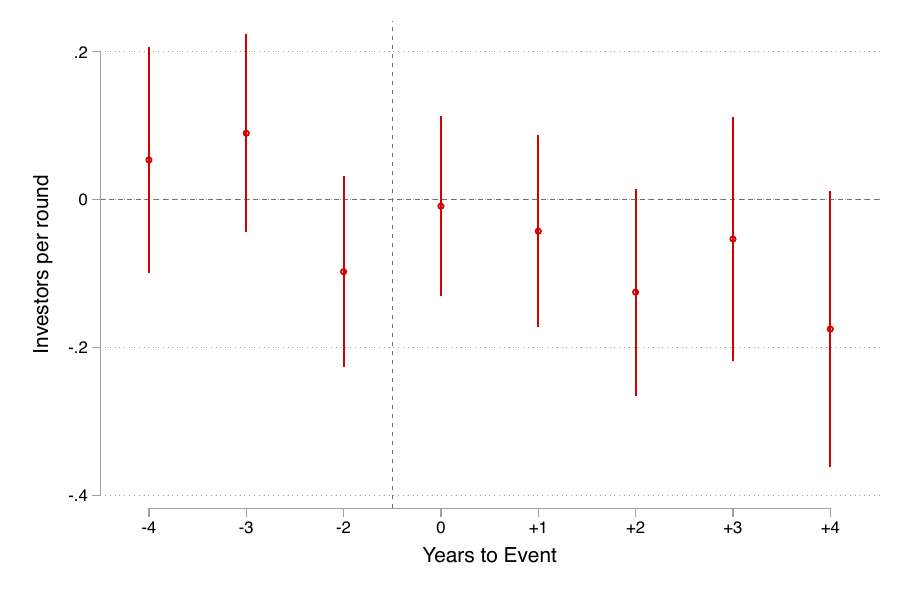}}
\caption{Dynamic Effect on the Investors per Round (Poisson)} \label{fig:investors_per_round_parallel}
\end{figure}

\begin{figure}[!htb]
\centerline{\includegraphics[width=5in]{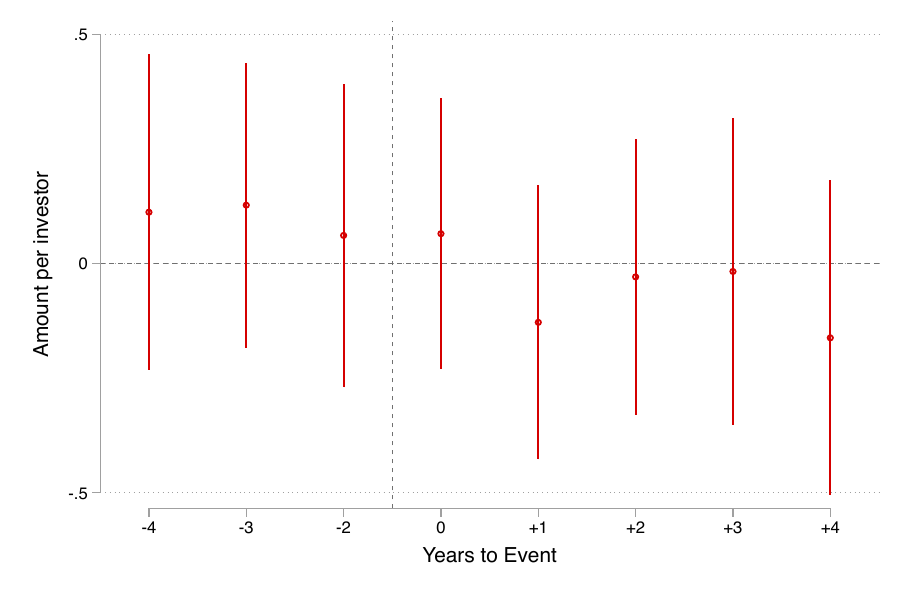}}
\caption{Dynamic Effect on the Amount per Investor (Poisson)} \label{fig:amount_per_investor_parallel}
\end{figure}

\begin{figure}[!htb]
\centerline{\includegraphics[width=5in]{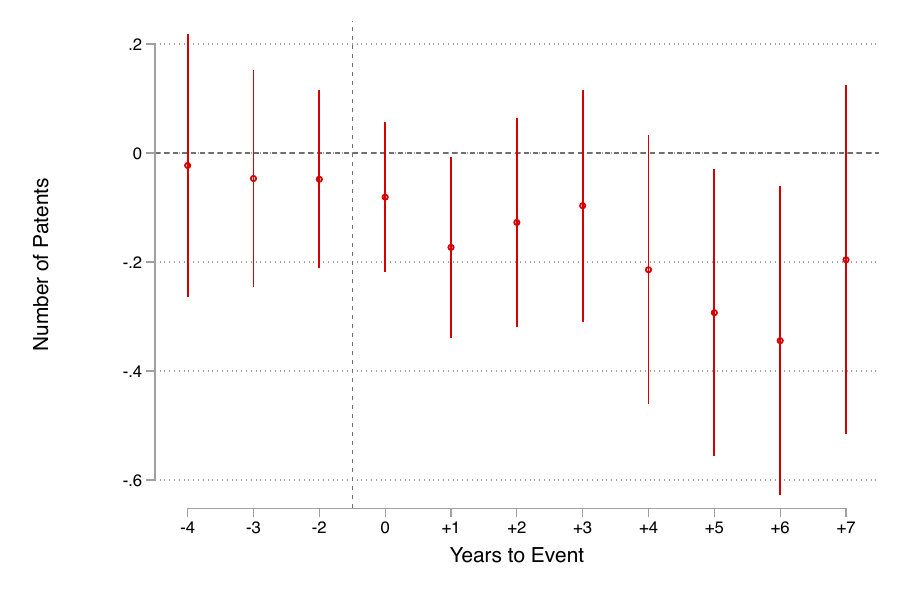}}
\caption{Dynamic Effect on the Number of Patents of VC-Backed Startups (Poisson)} 
\label{fig:numpat_parallel}
\end{figure}

To validate the parallel trend assumption of DID method, I estimate the dynamic effects of antitrust field office closures. The results are illustrated in Figure \ref{fig:vc_tot_amount_ols_parallel} - \ref{fig:numpat_parallel}. The coefficients of the difference between treated and control firms in all pre-event periods are not significantly different from that in 1 year before the office closures event, indicating the trends of treated and control firms are parallel before the event occurs.

\subsection{Tests for Concurrent Events}
To examine if any concurrent events happen to affect the economic landscape of treated states and drive the results, I use the specification in equation (2) to test if the annual GDP of treated states changes around the office closures event compared to the control states. As shown in Table \ref{table:robust_gdp}, the coefficients on the interaction term are insignificant, indicating the trajectory of the real GDP and real GDP per capita of treated states is not significantly different from that of control states around the event year. These results support that there are no concurrent events significantly affecting the economy of treated states.

\section{Conclusion}
This paper studies the effect of antitrust enforcement on VC investments and VC-backed companies. I employ difference-in-differences method with the antitrust field office closures in 2013 as an exogenous shock to the antitrust enforcement intensity. The results show that VCs significantly reduce investments in startups located in areas less protected by the antitrust division. This is true both on intensive margin and on extensive margin. Specifically, for startups already invested by VCs before the field office closures, the total amount of VC disbursement for startups in areas less policed by the antitrust division decreases significantly. This result is driven by both the reduction in the amount invested per investor and the number of investors involved in each round of VC funding. On the extensive margin, the number of new VC-backed startups in affected states decreases significantly. Additional tests show that the lowered antitrust enforcement subsequently leads to a lower likelihood of successful exits and worse innovation outcomes for startups in the affected areas. Overall, these results suggest that VCs avoid investing in startups located in areas less protected by the antitrust division, hindering the long-term development of startups in affected areas.

I further explore the potential mechanisms driving the negative effect of lowered antitrust enforcement. First, I find that the negative effect is stronger for startups in more concentrated industries where anticompetitive practices by incumbents take place more often. This result implies that anticompetitive practices against startups can be a mechanism driving the negative effect of less antitrust enforcement. Second, I find that innovative startups are less harmed by the reduction in antitrust enforcement. This result indicates that VCs are more confident in innovative startups' ability to innovate and differentiate their products from the incumbents' so as to escape the competition and avoid being abused by the incumbents. 

The empirical results in this paper speak to competition policies and local law enforcement agencies. Local antitrust enforcement is important and effective in promoting competition and innovation by protecting startups from being abused by the market power of local incumbents. The results also imply that a strategy for startups to mitigate the harm from lowered antitrust protection is to innovate and differentiate their products.

\clearpage
\bibliographystyle{jfe}
\bibliography{Reference}

\clearpage
\section*{Tables}
\begin{table}[htbp]\centering
\def\sym#1{\ifmmode^{#1}\else\(^{#1}\)\fi}
\caption{Summary Statistics}
\label{table:summary_stats}
\begin{tabular}{l*{1}{cccccc}}
\hline\hline
            &   N   &        Mean&          S.D.&         Min&         Median&         Max\\
\hline
$Treat$ &   80211&      0.279&      0.448&       0&       0&     1 \\
$\Delta Distance$ (Miles) &   80211&      80.289&     187.384&       0&       0&     798.732\\
Total amount (Thousand \$) &   80211&    2538.258&    7775.749&       0&       0&   50000\\
Amount per investor (Thousand \$)&   80211&     917.759&    2711.951&       0&       0&   18000\\
Investors per round&   80211&       0.754&       1.498&       0&       0&       7\\
Num of patents  &   80211&       0.410&       1.352&       0&       0&       9\\
HHI       &   60040&       0.256&       0.203&       0.059&       0.193&       0.999 \\
\hline\hline
\end{tabular}

\vspace{1ex}
{\footnotesize \raggedright \textit{Notes.} This table reports the summary statistics for variables listed in the first column. $Treat$ is a dummy variable equal to 1 if the startup becomes farther away from antitrust field offices after office closures in 2013. $\Delta Distance$ is the increase in geographical distance from the headquarter of a startup to the closest antitrust field office before and after the office closures. Total amount is the total amount of VC disbursement for a startup in a year. Amount per investor is the average amount of VC investments across investors. Investors per round is the number of investors involved in each round of VC financing. Num of patents is the number of patents applied by a startup in a year. HHI measures the concentration of public firms in the startup's industry identified by the 4-digit SIC code. \par}
\end{table}

\clearpage
\begin{table}[htbp]\centering
\def\sym#1{\ifmmode^{#1}\else\(^{#1}\)\fi}
\caption{The Effect of DOJ Antitrust Office Closures on VC Investments}
\label{table:vc_inves}
\begin{tabular}{l*{4}{c}}
\hline\hline
            
            &\multicolumn{1}{c}{Total Amount}&\multicolumn{1}{c}{Total Amount}&\multicolumn{1}{c}{Num of Investors}&\multicolumn{1}{c}{Amount }\\
            &\multicolumn{1}{c}{(Thousand \$)}&\multicolumn{1}{c}{(Thousand \$)}&\multicolumn{1}{c}{per Round}&\multicolumn{1}{c}{per Investor}\\
            \cmidrule{2-5}
            &        OLS         &        Poisson         &        Poisson &        Poisson    \\
            &\multicolumn{1}{c}{(1)}&\multicolumn{1}{c}{(2)}&\multicolumn{1}{c}{(3)}&\multicolumn{1}{c}{(4)}\\

\hline
$Treat_i \times Post_t$  &    -683.169\sym{***}&      -0.186\sym{***}&      -0.080\sym{**} &      -0.145\sym{**} \\
            &   (130.858)         &     (0.062)         &     (0.039)         &     (0.066)         \\
\hline
Startup FE & Y & Y & Y & Y \\
Year FE & Y & Y & Y & Y \\
\hline
$R^2$           &       0.250         &     0.424        &      0.220       &       0.370      \\
Num of Obs.     &   80211         &   80211         &   80211         &   80211         \\
\hline\hline
\end{tabular}

\vspace{1ex}
{\footnotesize \raggedright \textit{Notes.} This table reports the OLS and Poisson regression results of the startup-year level model: $Y_{it} = \alpha + \beta Treat_{i} \times Post_t + \theta_i + \gamma_t + \epsilon_{it}$, where $Treat_i$ is a dummy variable equal to 1 if startup $i$ becomes farther away from antitrust field offices after office closures in 2013 and $Post_t$ indicates years after antitrust field office closures in 2013. In column (1) and (2), the dependent variable is the total amount of VC disbursement for startup $i$ in year $t$. In column (3), the dependent variable is the number of investors involved in each round of VC financing. In column (4), the dependent variable is the average amount of VC investments across investors. Startup fixed effects and year fixed effects are included in all regressions. Standard errors are clustered by startup and are reported in parentheses below the coefficient estimates. *, **, and *** denote significance at the 10\%, 5\%, and 1\% levels, respectively. \par}
\end{table}

\clearpage
\begin{table}[htbp]\centering
\def\sym#1{\ifmmode^{#1}\else\(^{#1}\)\fi}
\caption{The Effect of DOJ Antitrust Office Closures on VC Exits}
\label{table:vc_exit}
\begin{tabular}{l*{3}{c}}
\hline\hline
            
            &\multicolumn{1}{c}{Success}&\multicolumn{1}{c}{Acquisition}&\multicolumn{1}{c}{IPO}\\
            \cmidrule{2-4}
            &        OLS         &      OLS        &        OLS     \\
            &\multicolumn{1}{c}{(1)}&\multicolumn{1}{c}{(2)}&\multicolumn{1}{c}{(3)}\\
\hline
$Treat_i$       &      -0.021\sym{**} &      -0.005         &      -0.015\sym{***}\\
            &     (0.010)         &     (0.009)         &     (0.004)         \\
$log(CumAmount2012)$ &       0.012\sym{***}&       0.009\sym{***}&       0.004\sym{***}\\
            &     (0.001)         &     (0.001)         &     (0.000)         \\
$Company \, Age$ &       0.000         &       0.000         &      -0.000         \\
            &     (0.001)         &     (0.001)         &     (0.000)         \\
$log(Real \, GDP \, per \, Capita)$ &       0.022         &       0.002         &       0.021         \\
            &     (0.028)         &     (0.026)         &     (0.013)         \\
\hline
Industry FE & Y & Y & Y \\
\hline
$R^2$          &       0.052         &       0.049         &       0.073         \\
Number of Obs.  &    9111         &    9111         &    9111         \\
\hline\hline
\end{tabular}

\vspace{1ex}

{\footnotesize \raggedright \textit{Notes.} This table reports the OLS regression results of the startup level model: $Y_{i} = \alpha + \beta_1 Treat_{i} + \beta_2 Controls_i + \gamma_f + \epsilon_{i}$, where $Treat_i$ is a dummy variable equal to 1 if startup $i$ becomes farther away from antitrust field offices after office closures in 2013. In column (1), the dependent variable is a dummy variable equal to 1 if startup $i$ successfully exited after 2013, and 0 otherwise. In column (2), the dependent variable is a dummy variable equal to 1 if startup $i$ exited by acquisition after 2013. In column (3), the dependent variable is a dummy variable equal to 1 if startup $i$ exited by IPO after 2013. Control variables include the log of the cumulative amount of VC investments received by startup $i$ up to the year before office closures (year 2012), startup age in 2012, and the log of real GDP per capita in 2012 of the state where the startup is located. Industry fixed effects are included in all regressions. Standard errors are reported in parentheses below the coefficient estimates. *, **, and *** denote significance at the 10\%, 5\%, and 1\% levels, respectively. \par}
\end{table}

\clearpage
\begin{table}[htbp]\centering
\def\sym#1{\ifmmode^{#1}\else\(^{#1}\)\fi}
\caption{The Effect of DOJ Antitrust Office Closures on the Innovation of VC-Backed Startups}
\label{table:vc_numpat}
\begin{tabular}{l*{5}{c}}

\hline\hline
            &\multicolumn{1}{c}{ Patents\_t}&\multicolumn{1}{c}{Patents\_t+1}&\multicolumn{1}{c}{Patents\_t+2}&\multicolumn{1}{c}{Patents\_t+3}&\multicolumn{1}{c}{Patents\_t+4}\\
            \cmidrule{2-6}
            &        Poisson         &        Poisson         &        Poisson &        Poisson   &        Poisson  \\
            &\multicolumn{1}{c}{(1)}&\multicolumn{1}{c}{(2)}&\multicolumn{1}{c}{(3)}&\multicolumn{1}{c}{(4)}&\multicolumn{1}{c}{(5)}\\ 
\hline
$Treat_i \times Post_t$  &      -0.167\sym{**} &      -0.138\sym{*}  &      -0.113         &      -0.134\sym{*}  &      -0.189\sym{**} \\
            &     (0.082)         &     (0.083)         &     (0.085)         &     (0.080)         &     (0.084)         \\
\hline
Startup FE & Y & Y & Y & Y & Y \\
Year FE & Y & Y & Y & Y & Y \\
\hline
$R^2$      &        0.453     &        0.467     &        0.482      &        0.483      &      0.517       \\
Num of Obs.    &   37573         &   33284         &   28582        &   24065      &   19641     \\
\hline\hline
\end{tabular}

\vspace{1ex}
{\footnotesize \raggedright \textit{Notes.} This table reports the Poisson regression results of the startup-year level model: $Y_{it+n} = \alpha + \beta Treat_{i} \times Post_t + \theta_i + \gamma_t + \epsilon_{it}$, where $Treat_i$ is a dummy variable equal to 1 if startup $i$ becomes farther away from antitrust field offices after office closures in 2013 and $Post_t$ indicates years after antitrust field office closures in 2013. In column (1), the dependent variable is the number of patents applied by startup $i$ in year $t$. In column (2) through (5), the dependent variable is lagged by 1 - 4 years. Startup fixed effects and year fixed effects are included in all regressions. Standard errors are clustered by startup and are reported in parentheses below the coefficient estimates. *, **, and *** denote significance at the 10\%, 5\%, and 1\% levels, respectively. \par}
\end{table}

\clearpage
\begin{table}[htbp]\centering
\def\sym#1{\ifmmode^{#1}\else\(^{#1}\)\fi}
\caption{The Effect on the Number of New VC-Backed Companies}
\label{table:new_vc_backed_companies}
\begin{tabular}{l*{3}{c}}
\hline\hline

            &\multicolumn{3}{c}{Num of New}\\
            &\multicolumn{3}{c}{VC-Backed Companies}\\
            \cmidrule{2-4}
            &        Poisson         &      Poisson        &        Poisson     \\
            &\multicolumn{1}{c}{(1)}&\multicolumn{1}{c}{(2)}&\multicolumn{1}{c}{(3)}\\            

\hline
$Treat_s \times Post_t$  &      -0.125\sym{*}  &      -0.125\sym{*}  &      -0.107\sym{*}  \\
            &     (0.069)         &     (0.069)         &     (0.064)         \\
$Treat_s$   &      -0.685         &                     &                     \\
            &     (0.562)         &                     &                     \\
$Post_t$     &       0.406\sym{***}&                     &                     \\
            &     (0.041)         &                     &                     \\
$log(Real \, GDP \, per \, Capita)$ &                     &                     &       0.767         \\
            &                     &                     &     (0.672)         \\
\hline
State FE & N & Y & Y \\
Year FE & N & Y & Y \\
\hline
$R^2$       &     0.055        &        0.934      &       0.933      \\
Num of Obs.    &     510         &     510         &     500         \\
\hline\hline
\end{tabular}

\vspace{1ex}
{\footnotesize \raggedright \textit{Notes.} This table reports the Poisson regression results of the state-year level model: $Y_{st} = \alpha + \beta_1 Treat_{s} \times Post_{t} + \beta_2 Treat_{s} + \beta_3 Post_{t} + \epsilon_{st}$, where $Treat_{s}$ is a dummy variable equal to 1 if state $s$ was overseen by an antitrust field office closed in 2013 and $Post_t$ indicates years after antitrust field office closures in 2013. The dependent variable is the number of new VC-backed startups in state $s$ in year $t$. Column (2) controls for state fixed effects and year fixed effects. Column (3) additionally controls for the natural logarithm of one plus the real GDP per capita of state $s$ in year $t$. Standard errors are clustered by state and are reported in parentheses below the coefficient estimates. *, **, and *** denote significance at the 10\%, 5\%, and 1\% levels, respectively. \par}
\end{table}

\clearpage
\begin{table}[htbp]\centering
\scriptsize
\def\sym#1{\ifmmode^{#1}\else\(^{#1}\)\fi}
\caption{Mechanism: Product Market Anticompetitive Conduct by Incumbents in Concentrated Industries}
\label{table:vc_inves_hhi}
\begin{tabular}{l*{8}{c}}
\hline\hline
            &\multicolumn{2}{c}{Total Amount} &\multicolumn{2}{c}{Total Amount}&\multicolumn{2}{c}{Num of Investors}&\multicolumn{2}{c}{Amount}\\
            &\multicolumn{2}{c}{(Thousand \$)} &\multicolumn{2}{c}{(Thousand \$)}&\multicolumn{2}{c}{per Round}&\multicolumn{2}{c}{per Investor}\\
            &\multicolumn{2}{c}{OLS} &\multicolumn{2}{c}{Poisson}&\multicolumn{2}{c}{Poisson}&\multicolumn{2}{c}{Poisson}\\
\cmidrule{2-9}
Industry Concentration    &    High    &    Low    &    High    &    Low    &   High    &    Low   &   High    &    Low  \\
            &\multicolumn{1}{c}{(1)}&\multicolumn{1}{c}{(2)}&\multicolumn{1}{c}{(3)}&\multicolumn{1}{c}{(4)}&\multicolumn{1}{c}{(5)}&\multicolumn{1}{c}{(6)}&\multicolumn{1}{c}{(7)}&\multicolumn{1}{c}{(8)}\\

\hline
$Treat_i \times Post_t$  &   -1176.110\sym{***}&     375.179         &      -0.250\sym{***}&       0.077         &      -0.184\sym{***}&       0.083         &      -0.202\sym{**} &       0.136         \\
            &   (175.702)         &   (311.661)         &     (0.078)         &     (0.148)         &     (0.049)         &     (0.093)         &     (0.085)         &     (0.154)         \\
\hline
Startup FE & Y & Y & Y & Y & Y & Y & Y & Y \\
Year FE & Y & Y & Y & Y & Y & Y & Y & Y \\
\hline
$R^2$       &       0.258         &       0.244         &       0.428       &         0.420       &      0.210          &        0.255       &       0.365      &     0.385           \\
Num of Obs.       &   44640         &   15374         &   44640        &   15374        &   44640         &   15374         &   44640      &   15374        \\
\hline\hline
\end{tabular}

\vspace{1ex}
{\footnotesize \raggedright \textit{Notes.} This table reports the OLS and Poisson regression results of the startup-year level model: $Y_{it} = \alpha + \beta Treat_{i} \times Post_t + \theta_i + \gamma_t + \epsilon_{it}$, where $Treat_i$ is a dummy variable equal to 1 if startup $i$ becomes farther away from antitrust field offices after office closures in 2013 and $Post_t$ indicates years after antitrust field office closures in 2013. In columns (1) - (4), the dependent variable is the total amount of VC disbursement for startup $i$ in year $t$. In columns (5) - (6), the dependent variable is the number of investors involved in each round of VC financing. In columns (7) - (8), the dependent variable is the average amount of VC investments across investors. Column (1), (3), (5) and (7) report the estimation results using the subsample where startups are in more concentrated industries. Column (2), (4), (6) and (8) report the estimation results using the subsample where startups are in more competitive industries. Startup fixed effects and year fixed effects are included in all regressions. Standard errors are clustered by startup and are reported in parentheses below the coefficient estimates. *, **, and *** denote significance at the 10\%, 5\%, and 1\% levels, respectively. \par}
\end{table}

\clearpage
\begin{table}[htbp]\centering
\scriptsize
\def\sym#1{\ifmmode^{#1}\else\(^{#1}\)\fi}
\caption{Mechanism: Potential to be a Future Monopolist}
\label{table:vc_inves_innovative}
\begin{tabular}{l*{8}{c}}
\hline\hline
            &\multicolumn{2}{c}{Total Amount} &\multicolumn{2}{c}{Total Amount}&\multicolumn{2}{c}{Num of Investors}&\multicolumn{2}{c}{Amount}\\
            &\multicolumn{2}{c}{(Thousand \$)} &\multicolumn{2}{c}{(Thousand \$)}&\multicolumn{2}{c}{per Round}&\multicolumn{2}{c}{per Investor}\\
            &\multicolumn{2}{c}{OLS} &\multicolumn{2}{c}{Poisson}&\multicolumn{2}{c}{Poisson}&\multicolumn{2}{c}{Poisson}\\
\cmidrule{2-9}
Average num of patents    &    High    &    Low    &    High    &    Low    &   High    &    Low   &   High    &    Low  \\
per year pre-event            &\multicolumn{1}{c}{(1)}&\multicolumn{1}{c}{(2)}&\multicolumn{1}{c}{(3)}&\multicolumn{1}{c}{(4)}&\multicolumn{1}{c}{(5)}&\multicolumn{1}{c}{(6)}&\multicolumn{1}{c}{(7)}&\multicolumn{1}{c}{(8)}\\
\hline
$Treat_i \times Post_t$  &    -608.115\sym{*}  &    -712.221\sym{***}&      -0.118         &      -0.239\sym{***}&      -0.092         &      -0.071         &       0.059         &      -0.181\sym{***}\\
            &   (345.164)         &   (134.384)         &     (0.105)         &     (0.078)         &     (0.070)         &     (0.047)         &     (0.092)         &     (0.069)         \\
\hline
Startup FE & Y & Y & Y & Y & Y & Y & Y & Y \\
Year FE & Y & Y & Y & Y & Y & Y & Y & Y \\
\hline
$R^2$         &       0.235         &       0.250         &       0.367       &      0.436       &       0.209        &       0.216       &        0.310       &    0.367           \\
Num of Obs.      &   19036         &   61175         &   19036         &   61175         &   19036        &   61175        &   19036      &   61175      \\
\hline\hline
\end{tabular}

\vspace{1ex}
{\footnotesize \raggedright \textit{Notes.} This table reports the OLS and Poisson regression results of the startup-year level model: $Y_{it} = \alpha + \beta Treat_{i} \times Post_t + \theta_i + \gamma_t + \epsilon_{it}$, where $Treat_i$ is a dummy variable equal to 1 if startup $i$ becomes farther away from antitrust field offices after office closures in 2013 and $Post_t$ indicates years after antitrust field office closures in 2013. In columns (1) - (4), the dependent variable is the total amount of VC disbursement for startup $i$ in year $t$. In columns (5) - (6), the dependent variable is the number of investors involved in each round of VC financing. In columns (7) - (8), the dependent variable is the average amount of VC investments across investors. Column (1), (3), (5) and (7) report the estimation results using the subsample where the startups' average number of patents generated per year before the event is above the median. Column (2), (4), (6) and (8) report the estimation results using the subsample where the startups' average number of patents generated per year before the event is below the median. Startup fixed effects and year fixed effects are included in all regressions. Standard errors are clustered by startup and are reported in parentheses below the coefficient estimates. *, **, and *** denote significance at the 10\%, 5\%, and 1\% levels, respectively. \par}
\end{table}

\clearpage
\begin{table}[htbp]\centering
\def\sym#1{\ifmmode^{#1}\else\(^{#1}\)\fi}
\caption{Robustness Test: GDP Change around the Event}
\label{table:robust_gdp}
\begin{tabular}{l*{2}{c}}
\hline\hline

            &\multicolumn{1}{c}{log(1+Real GDP)}&\multicolumn{1}{c}{log(1+Real GDP per Capita)}\\
            &\multicolumn{1}{c}{(1)}&\multicolumn{1}{c}{(2)}\\
\hline
$Treat_s \times Post_t$  &      -0.013         &      -0.006         \\
            &     (0.015)         &     (0.013)         \\
\hline
State FE & Y & Y \\
Year FE & Y & Y \\
\hline
$R^2$         &       0.999         &       0.985         \\
Num of Obs.           &     510         &     510         \\
\hline\hline
\end{tabular}

\vspace{1ex}
{\footnotesize \raggedright \textit{Notes.} This table reports the OLS regression results of the state-year level model: $Y_{st} = \alpha + \beta_1 Treat_{s} \times Post_{t} + \beta_2 Treat_{s} + \beta_3 Post_{t} + \epsilon_{st}$, where $Treat_{s}$ is a dummy variable equal to 1 if state $s$ was overseen by an antitrust field office closed in 2013 and $Post_t$ indicates years after antitrust field office closures in 2013. In column (1), the dependent variable is the natural logarithm of one plus the real GDP of state $s$ in year $t$. In column (2), the dependent variable is the natural logarithm of one plus the real GDP per capita of state $s$ in year $t$. State fixed effects and year fixed effects are included in all regressions. Standard errors are clustered by state and are reported in parentheses below the coefficient estimates. *, **, and *** denote significance at the 10\%, 5\%, and 1\% levels, respectively. \par}
\end{table}

\clearpage
\section*{Figures}
\vfill
\vfill

\vfill
\vfill

\vfill
\vfill

\vfill
\vfill

\vfill
\vfill



\end{document}